\documentclass[pra,twocolumn,showpacs,preprintnumbers,amsmath,amssymb,tightenlines,epsfig,superscriptaddress]{revtex4}
\usepackage{amsmath}
\usepackage{amssymb}
\usepackage{amsthm}
\usepackage{amsfonts}
\usepackage{graphicx}
\usepackage{bm}
\usepackage{xcolor}
\usepackage[percent]{overpic}
\usepackage{bbm}
\usepackage{lipsum}
\usepackage[normalem]{ulem}

\newcommand{\bra}[1]{\langle\,{#1}\, |}
\newcommand{\ket}[1]{|\,{#1}\,\rangle}
\newcommand{\braket}[2]{\mbox{$\langle\,{#1}\, | \,{#2}\,\rangle$}}

\newcommand{\xvec}{\mathbf{x}}
\newcommand{\qvec}{\mathbf{q}}

\newcommand{\rvec}[1]{{\bv{r}}}

\newcommand{\cohst}{\boldsymbol{d}}
\newcommand{\cohstval}{d}
\newcommand{\psihops}{f}

\newcommand{\ssection}[1]{{\noi  \it #1:}}

\newcommand{\expec}[1]{\langle #1 \rangle}

\newcommand{\sub}[2]{{#1}_{\mbox{\!\! \scriptsize #2}}}

\newcommand{\bv}[1]{\mathbf{ #1 }}

\newcommand{\xv}{\mathbf{x}}

\def\noi{\noindent}
\def\beq{\begin{equation}}
\def\eeq{\end{equation}}

\def\CR{\nonumber\\[0.15cm]}

\newcommand{\rref}[1]{Ref.~\cite{#1}}
\newcommand{\fref}[1]{Fig.~\ref{#1}}
\newcommand{\frefp}[2]{Fig.~\ref{#1}~(#2)}

\newcommand{\eref}[1]{Eq.~(\ref{#1})}

\newcommand{\cref}[1]{chapter~\ref{#1}}

\newcommand{\Cref}[1]{Chapter~\ref{#1}}

\newcommand{\bref}[1]{(\ref{#1})}

\usepackage{ulem}  
\begin{document}

\title{Imaging the interface of a qubit and its quantum-many-body environment}
\author{S.~Rammohan}
\affiliation{Department of Physics, Indian Institute of Science Education and Research, Bhopal, Madhya Pradesh 462 066, India}
\author{S.~Tiwari}
\affiliation{Department of Physics, Indian Institute of Science Education and Research, Bhopal, Madhya Pradesh 462 066, India}
\author{A.~Mishra}
\affiliation{Department of Physics, Indian Institute of Science Education and Research, Bhopal, Madhya Pradesh 462 066, India}
\author{A.~Pendse}
\affiliation{Department of Physics, Indian Institute of Science Education and Research, Bhopal, Madhya Pradesh 462 066, India}
\author{A.~K.~Chauhan}
\affiliation{Department of Physics, Indian Institute of Science Education and Research, Bhopal, Madhya Pradesh 462 066, India}
\affiliation {Department of Optics, Faculty of Science, Palack\'y University, 17.~listopadu 1192/12, 77146 Olomouc, Czech Republic}
\author{R.~Nath}
\affiliation{Department of Physics, Indian Institute of Science Education and Research, Pune 411 008, India}
\author{A.~Eisfeld}
\affiliation{Max Planck Institute for the Physics of Complex Systems, N\"othnitzer Strasse 38, 01187 Dresden, Germany}
\author{S.~W\"uster}
\affiliation{Department of Physics, Indian Institute of Science Education and Research, Bhopal, Madhya Pradesh 462 066, India}
\email{sidharth16@iiserb.ac.in, sebastian@iiserb.ac.in}
\begin{abstract}
Decoherence affects all quantum systems, natural or artificial, and is the primary obstacle impeding quantum technologies.
We show theoretically that for a Rydberg qubit in a Bose condensed environment, experiments can image the system-environment interface that is central for decoherence.
High precision absorption images of the condensed environment will be able to capture transient signals
that show the real time build up of a mesoscopic entangled state in the environment. This is possible before decoherence sources other than the condensate itself can kick in, since qubit decoherence time-scales can be tuned from the order of nanoseconds to microseconds by choice of the excited Rydberg principal quantum number $\nu$.
Imaging the interface will allow detailed explorations of open quantum system concepts and may offer guidance for coherence protection in challenging scenarios with non-Markovian environments.
\end{abstract}

\maketitle

\ssection{Introduction} 
Quantum decoherence \cite{schloss_decoherence,Schlosshauer_decoherence_review}  is central to quantum science and technologies, for which it mainly represents an obstacle but also is increasingly considered a possible resource \cite{Poyatos_dissip_engineering_PhysRevLett,Verstraete_dissip_engineering,Vuglar_nonconsforces_PhysRevLett}. Decoherence also often reconciles our every-day experience with the laws of quantum-mechanics, explaining the gradual loss of interference features once a hitherto isolated quantum object begins to interact with its environment. Disconcertingly, the cause is not a reluctance of the macroscopic environment to participate in non-classical aspects of quantum physics such as superposition states, but its propensity to entangle with every microscopic system interacting with it.

This entanglement diminishes interference features involving the quantum object, to the point where they become un-observable. 
Experiments are typically confined to measurements of the quantum object, since it is difficult to extract useful information about a large and complex environment.
\begin{figure}[htb]
\includegraphics[width=0.99\columnwidth]{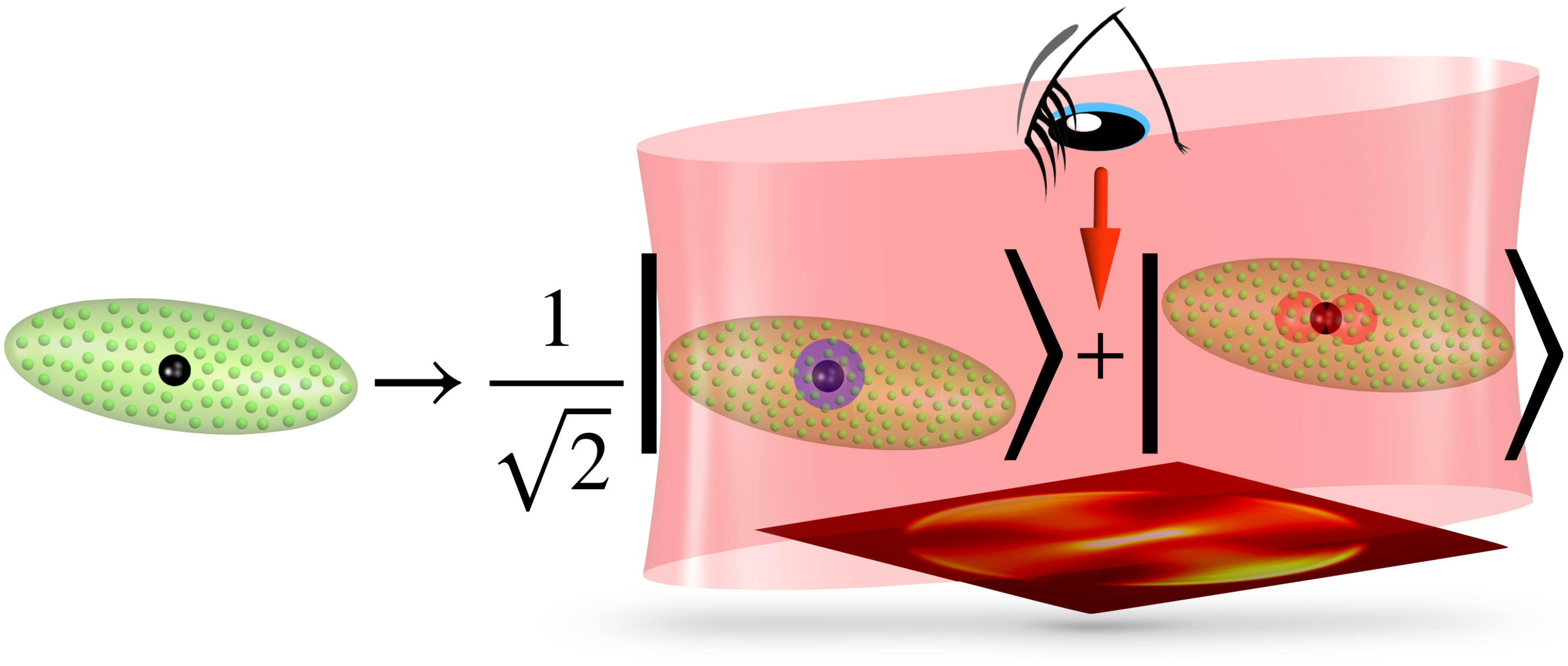}
\caption{\label{sketch} 
Imaging the environmental origin of decoherence.
 (left) We assume a distinct impurity atom (dark spot) is embedded at the centre of the BEC (green ellipsoid) of ground-state atoms (green balls).
(right) The impurity is rapidly transferred into a coherent superposition of two different electronic Rydberg states, $\ket{s}$ (blue shade) and $\ket{p}$ (orange shade) forming a qubit. In either state the
 electron makes large excursions into the ambient BEC medium, within the orbital volume indicated by the blue/orange shaded areas. In this volume, the electron will imprint a phase pattern onto the coherent condensate wave-function, entangling it with the qubit. Due to the well controlled and coherent environmental initial state, this entangled pattern can be imaged (bottom) with high sensitivity absorption imaging (red laser beam).
}
\end{figure}
Here we propose an exceptional platform where crucial parts of the environment around a qubit can also be probed.
We show that exciting an impurity atom in a Bose-Einstein-Condensate (BEC) to Rydberg states \cite{heidemann2008rydberg,balewski:elecBEC,gaj:molspecdensshift,schlagmuller2016probing,camargo2018creation,Dieterle_inelasticions_BEC_PRA,middelkamp:rydinBEC,Whalen_Rydbmol_lifetimes_BEC,Kanungo_Sr_lossrates}  provides a scenario where both, the reduced dynamics of the Rydberg \cite{book:gallagher,loew:rydguide:jpbreview} qubit and the evolution of its environment are observable. Even the build-up of mesoscopic entanglement during the decoherence time can be probed.

This direct window on the interface between a qubit and its environment is enabled by the extraordinary properties of both, Rydberg atoms and BEC. Firstly, the atom can simultaneously interact with a large portion of the environment due to the wide excursions of its electron \cite{schlagmuller2016probing}. Secondly, the many-body quantum state of the BEC environment can be well approximated by a quite simple product ansatz \cite{book:pethik} that nonetheless describes a large number of atoms. This allows the observation of the BEC evolution while decoherence of the Rydberg system takes place. 
Probing the environment while it is causing decoherence would be significantly more challenging around ground-state qubits, on which  
earlier work regarding the decoherence of impurity atoms within a BEC has focussed \cite{schmidt2018quantum,schmidt2019motional,song2019controlling,ostmann2017cooling,mcendoo2013entanglement,klein2007dynamics,cirone2009collective,bruderer2007polaron,bruderer2008transport,bruderer2006probing,haikka2013non}.

The qubit here is based on the pseudo spin-1/2 of a Rydberg atom with large principal quantum number (e.g.~$\nu=80$) created in a BEC of $N$ atoms, and rapidly brought into a quantum superposition $\ket{\chi} =c_\downarrow \ket{s} + c_\uparrow \ket{p}$ of two different angular momentum states $\ket{s}=\ket{\nu (l=0)}\equiv\ket{\downarrow}$ and $\ket{p}=\ket{\nu (l=1,m=0)}\equiv\ket{\uparrow}$. In either state, hundreds of condensate atoms are located within the Rydberg electron orbit, see \fref{sketch}. Interactions of these atoms with the electron then imprint a macroscopic phase pattern onto the condensate \cite{mukherjee:phaseimp}. Crucially, this pattern is different for different electronic states \cite{Karpiuk_imaging_NJP}, hence the condensate environment entangles with the Rydberg impurity qubit, causing decoherence of the latter. 

We will demonstrate that the proposed platform is exceptionally well suited to interrogate the intertwined dynamics of the qubit and its environment that lead to deocherence: (i) qubit decoherence is accessible by Ramsey microwave interferometry \cite{Mukherjee_Ramsey_PhysRevA,Arias_dressed_Ramsey,Dietsche_Ramsey_circular}. (ii) Absorption images show the transient build-up of the mesoscopically entangled state causing decoherence. (iii) One can then push the platform towards highly nontrivial, non-Markovian quantum many-body dynamics, where observations will challenge the best numerical techniques available. Such studies  can provide new insight into avoiding decoherence or exploiting it for quantum technologies \cite{Poyatos_dissip_engineering_PhysRevLett,Verstraete_dissip_engineering,Vuglar_nonconsforces_PhysRevLett}. 

\ssection{Interactions between qubit and environment} 
%
We have shown in \cite{rammohan_tayloring}, starting from the many-body Hamiltonian for the two species system, that electronic dynamics of the Rydberg impurity embedded in the BEC is described in the Bogoliubov approximation by the Spin-Boson model \cite{breuer2002theory,Leggett_dissipative_twolevel,wilson2002quantum,xu1994coupling,leppakangas2018quantum} (SBM) $\sub{\hat{H}}{tot}=\sub{\hat{H}}{syst}+\sub{\hat{H}}{env}+\sub{\hat{H}}{int}$, in terms of the Rydberg pseudo-spin defined earlier, with 
\begin{align}
\sub{\hat{H}}{syst}& = \frac{\sub{\Omega}{mw}}{2} \hat{\sigma}_x +  \frac{\Delta E(t)}{2} \hat{\sigma}_z, \:\:\:\sub{\hat{H}}{env}= \sum_\qvec\hbar\omega_\qvec \:\tilde{b}^{\dagger}_\qvec\tilde{b}_\qvec,\CR
\sub{\hat{H}}{int}&=\sum_\mathbf{q}\frac{\Delta\kappa_\mathbf{q}}{2}\Big(\tilde{b}_\mathbf{q}+\tilde{b}_\mathbf{q}^{\dagger}\Big)\hat{\sigma}_z.
\label{Hspinboson}
\end{align}
The $\hat{\sigma}_k$ are Pauli operators and the operators $\tilde{b}_\mathbf{q}$ destroy Bogoliubov-de-Gennes (BdG) excitations with wave vector $\qvec$. Further, $\sub{\Omega}{mw}$ is the Rabi frequency of the micro-wave coupling between the $\ket{p}$ and $\ket{s}$ Rydberg state, with $\Delta E(t)=\Delta + \bar{E}(t)$, where $\Delta$ is the effective energy splitting
without the small time-dependent shift $\bar{E}(t)$ due to the environment 
\cite{rammohan_tayloring}. The BdG mode energies are $\hbar\omega_q=\sqrt{E_q(E_q + U_0 \rho)}$, with $q=|\qvec|$ and $E_q=\hbar^2 q^2/2m$ for a homogeneous condensate of atoms with mass $m$ at density $\rho$ and interaction strength $U_0$. Finally, the coupling-strengths to BdG modes are 
\cite{rammohan_tayloring}
\begin{align} 
\label{kappa}
\Delta\kappa_\mathbf{q} &=g_0 \sqrt{\rho}  \int d^3\xv  (|\psi^{(p)}({\xvec})|^2 - |\psi^{(s)}({\xvec})|^2 ) [u_\qvec ({\xvec})-v^*_\qvec ({\xvec})], 
\end{align} 
where $g_0$ is the strength of electron-atom interactions 
\cite{omont1977theory,footnote:GPEparameters}, $\psi^{(\alpha)}(\xv)$ the Rydberg electron wave-function in state $\alpha$ and
$u_{\qvec}(\xv)=\frac{\bar{u}_\qvec}{\sqrt{\cal V}} e^{i\qvec \xvec}$, $v_{\qvec}(\xv)=\frac{\bar{v}_\qvec}{\sqrt{\cal V}} e^{i\qvec \xvec}$ 
BdG mode functions. ${\cal V}$ is the quantisation volume, and $\bar{u}_\qvec$, $\bar{v}_\qvec$ BdG amplitudes \cite{book:pethik}.

\ssection{Qubit decoherence} 
%
As a prelude to interface imaging in the next section, we now first focus on decoherence of the qubit itself.
This already allows us to predict decoherence time-scales from a very simple expression. These will turn out highly tunable, which will be essential for practical implementation.
To estimate the decoherence time-scale, it is sufficient to consider the simple scenario when there is no coupling between the two spin states, implying $\sub{\Omega}{mw}=0$. 

Thus, while we assume a micro-wave is adiabatically followed to create the superposition qubit state $\ket{\chi}$ with $c_\downarrow=c_\uparrow=1/\sqrt{2}$, which we call $\ket{+}$, the microwave should then subsequently be switched out. We can hence study the Hamiltonian $\sub{\hat{H}}{tot}$ with $\sub{\Omega}{mw}=0$ and $\Delta=E_p-E_s$, the energy difference between the Rydberg states. The many-body Hamiltonian then de-composes into blocks $\sub{\hat{H}}{tot}=\ket{\uparrow}\bra{\uparrow}\otimes\hat{H}_{ph,\uparrow}  +\ket{\downarrow}\bra{\downarrow}\otimes\hat{H}_{ph,\downarrow} $, where $\hat{H}_{ph,\downarrow/\uparrow}$ pertain to the environment (phonon) space only. Hence also the time-evolution can be separately found in each of those blocks, yielding a global time-evolution operator $\hat{U}= \ket{\uparrow}\bra{\uparrow}\otimes\hat{U}_{ph,\uparrow}+ \ket{\downarrow}\bra{\downarrow}\otimes\hat{U}_{ph,\downarrow}$, where $\hat{U}_{ph,\uparrow}$ ($\hat{U}_{ph,\downarrow}$) denotes a unitary operator that acts on BdG modes only, for the case that the system is in the $\ket{\uparrow}$ ($\ket{\downarrow}$) state. The above $\hat{U}$ implies a time-evolved state
\begin{align}
\ket{\sub{\Psi}{tot}(t)}&=c_\uparrow \ket{\uparrow}\otimes \ket{\Psi_\uparrow(t)} + c_\downarrow \ket{\downarrow}\otimes \ket{\Psi_\downarrow(t)}.
\label{psitot}
\end{align}
Here and in the following, states labelled $\Psi$ are quantum many-body states. In \bref{psitot}, $\ket{\sub{\Psi}{tot}(t)}$ includes the qubit while $\ket{\Psi_{\uparrow,\downarrow}(t)}$ describe the environment. For a state of the bipartite form \bref{psitot}, the coherence between $\ket{\uparrow}$ and $\ket{\downarrow}$ in the reduced density matrix for the qubit is \cite{schloss_decoherence,Schlosshauer_decoherence_review}
\begin{align}
\rho_{\downarrow\uparrow}(t) &= c^*_\downarrow c_\uparrow \braket{\Psi_\uparrow(t)}{\Psi_\downarrow(t)}.
\label{cohfact}
\end{align}
We will refer to $|r(t)|=|\braket{\Psi_\uparrow(t)}{\Psi_\downarrow(t)}|$ as the coherence factor. It can be evaluated explicitly from the total  time evolution operator $\hat{U}(t)={\cal T}\exp\Big[-\frac{i}{\hbar}\int_0^t\sub{\hat{H}}{int}'(t')\:dt'\Big]$ of the SBM, where $\sub{\hat{H}}{int}'$ denotes $\sub{\hat{H}}{int}$ in the interaction picture. The two blocks discussed above are then
\begin{align}
\label{U_s_unitary_ope}
\hat{U}_{ph,\alpha}(t)&=\exp\Bigg[\alpha\sum_\qvec\frac{\Delta\kappa_\qvec}{\hbar\omega_\qvec}\Bigg(\tilde{b}_\qvec(0)(e^{-i\omega_\qvec t}-1) \CR
& -\tilde{b}_\qvec^{\dagger}(0)(e^{i\omega_\qvec t}-1)\Bigg)\Bigg],
\end{align}
with $\alpha\in\{-\frac{1}{2},\frac{1}{2}\}\leftrightarrow \{\downarrow,\uparrow \}$. 
In the Bogoliubov vacuum we then find
\begin{align}
\label{decoh_factor2}
r(t)=\bra{0}\hat{U}_{ph,\uparrow}^{\dagger}(t)\hat{U}_{ph,\downarrow}(t)\ket{0},
\end{align}
which yields
\begin{align}
\label{r_t_final}
r(t)&=\text{exp}\Bigg[-\sum_\qvec\Bigg(\frac{\Delta\kappa_\qvec}{\hbar\omega_\qvec}\Bigg)^2(1-\cos(\omega_\qvec t))\Bigg].
\end{align}
%
For short times, we can approximate
\begin{align}
|r(t)|&\approx e^{-\left(t/\sub{T}{dc}\right)^2}, \:\sub{T}{dc}=\Big(\sum_\qvec \Delta\kappa_\qvec^2/(2\hbar)\Big)^{-1/2},
\label{decoh_timescale}
\end{align}
with decoherence time-scale $\sub{T}{dc}$. Exploiting the framework set up in \cite{rammohan_tayloring}, we find from \bref{decoh_timescale} that $\sub{T}{dc}$ can be tuned from $\sub{T}{dc}\approx 20$ ns at $\nu =40$ to  $\sub{T}{dc}\approx 1$ $\mu$s at $\nu =120$ through choice of principal quantum number $\nu$ and other parameters given in \cite{footnote:GPEparameters}. 
The coherence evolution $|r(t)|$ of the Rydberg impurity is directly measurable with micro-wave Ramsey interferometry \cite{Mukherjee_Ramsey_PhysRevA,Arias_dressed_Ramsey,Dietsche_Ramsey_circular}. For most of the range of $\sub{T}{dc}$ above, the process described will be faster than other decoherence sources, such as black-body radiation \cite{beterov:bbr}, phonon-phonon interactions \cite{Howl_phonondecoh_JPB} or atomic losses \cite{jack:threebodyloss,dziarmaga:lossheating,pendse:decoherence}, as well as radiative lifetimes \cite{footnote:lifetimes}. Since \eref{decoh_timescale} depends on the microscopic interactions between Rydberg and BEC atoms, decoherence measurements will open an additional route to understand these.

\ssection{Extracting decoherence from environmental mean-field dynamics}
%
Spin coherence is intimately linked to the dynamics of the environment through \eref{cohfact}. We now show that the extreme coherence of the BEC environment allows experimental imaging of this joint dynamics, as well as the prediction of the expected signature in a simple case. Based on the block-decomposition of the many-body Hamiltonian for $\sub{\Omega}{mw}=0$, we know that for each block separately, the BEC evolution can be described using the Gross-Pitaevskii equation (GPE) \cite{tiwari:tracking}. Decoherence of the superposition is then captured by a combination of these individual condensate mean-field solutions. 
We thus write the many-body wavefunction for the Bose gas as $\braket{\mathbf{X}}{\sub{\Psi}{$\alpha$}(t)}=\prod_k^N \varphi_\alpha(\mathbf{x}_k)$, for $\alpha\in\{\uparrow,\downarrow\}$, where $\mathbf{X}\in {\mathbb{R}}^{3N}$ groups all atomic positions, while $\mathbf{x}_k$ are those of atom $k$ only. Thus for a fixed Rydberg state $\alpha$, all ground-state atoms are in the same single particle state $\ket{\varphi_\alpha}$, while through \eref{psitot} this single particle state is entangled with the impurity state.

\begin{figure}[htb]
\includegraphics[width=0.99\columnwidth]{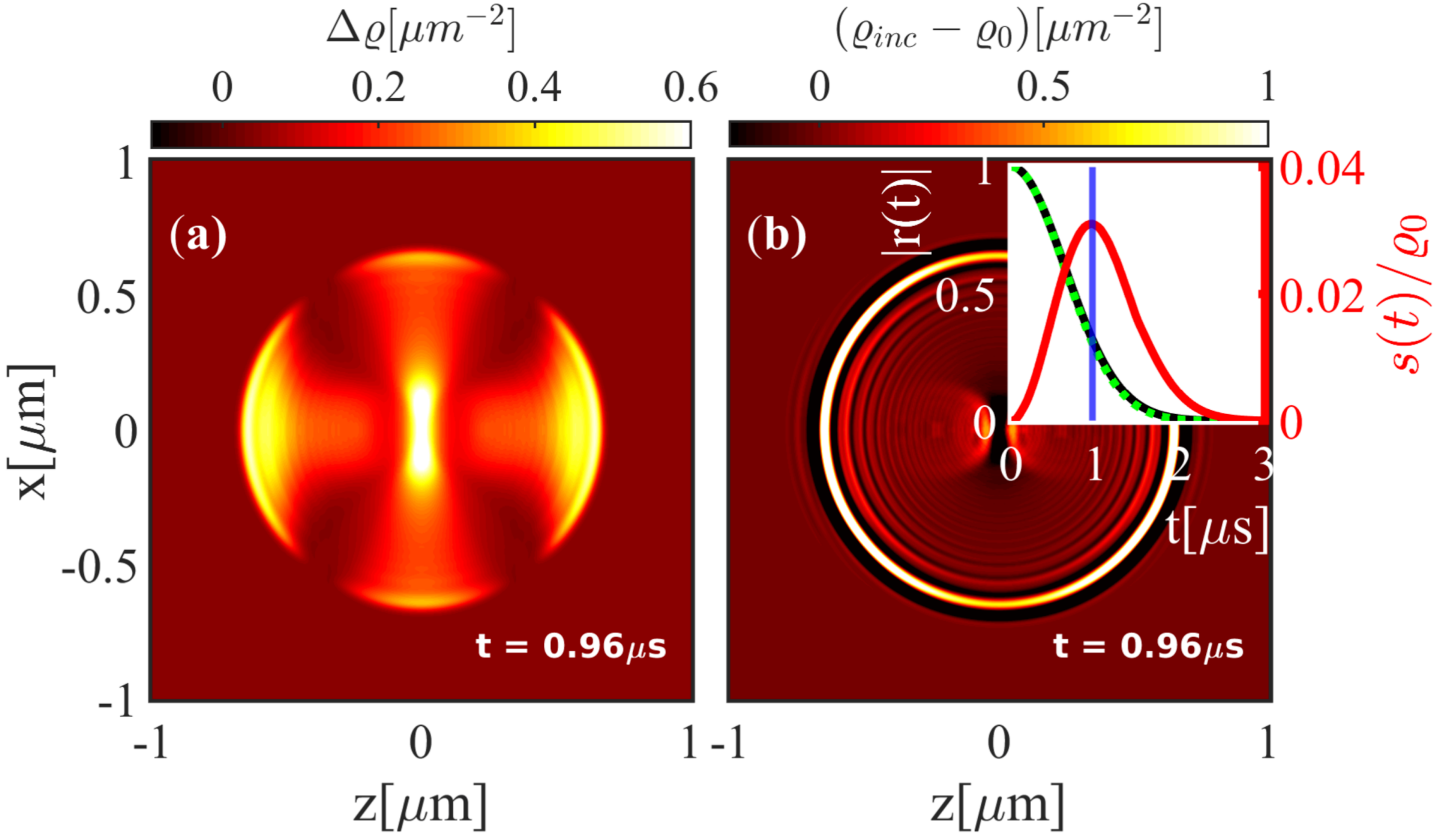}
\caption{\label{coherence_factor} Interface imaging. (a) The difference $\Delta\varrho=\sub{\varrho}{ms}(\bv{r})-\sub{\varrho}{inc}(\bv{r})$ in \eref{dens_diff} of BEC column densities for two scenarios, where $\sub{\varrho}{inc}=(\varrho_\uparrow + \varrho_\downarrow)/2$ is a random mixture of the patterns imprinted by a $\ket{s}=\ket{\downarrow}$ or $\ket{p}=\ket{\uparrow}$ Rydberg state at $\nu=80$, while for $\sub{\varrho}{ms}$ the BEC is in a mesoscopic superposition of these two patterns. See also supplementary movie \cite{sup:info}. (b) As reference we show $\sub{\varrho}{inc}$ alone, relative to the background value
 $\varrho_0=24.1{\mu}m^{-2}$. The inset shows the Rydberg coherence factor $|r(t)|$ between $\ket{s}$ and $\ket{p}$, with $|r(t)|$ from the SBM in \eref{decoh_timescale} (green dotted), from GPE orbital overlap in \eref{cohfact_from_GPE} (black) and the spatial maximum $s(t)=\max_r\Delta\varrho{(r,t)}$ (red solid line) peaking at $t=0.128$ $\mu$s used in (a,b) (blue vertical line).}
\end{figure}
 If there is an impurity in the state $\ket{\alpha}$ located at $\bv{r}=0$, the three-dimensional (3D) GPE reads \cite{middelkamp:rydinBEC,mukherjee:phaseimp,tiwari:tracking,Karpiuk_imaging_NJP,Shukla_pandit_particlesinsuperfluids}:
\begin{align}
&i\hbar \frac{\partial}{\partial t}\phi_{\alpha}(\bv{r})= \bigg(-\frac{\hbar^2}{2 m}\boldsymbol{\nabla}^2  +U_0 |\phi_{\alpha}(\bv{r})|^2 + g_0 |\psi^{(\alpha)}(\bv{r})|^2 \bigg)\phi_{\alpha}(\bv{r}),
\label{impurityGPE}
\end{align}
where $\psi^{(\alpha)}$ is the Rydberg electron wave-function of the impurity and $\phi_\alpha= \sqrt{N}\varphi_\alpha$ the BEC mean-field wavefunction.

With the separate mean-field ansatz for each impurity state discussed above, the coherence factor becomes
\begin{align}
|\sub{r}{GPE}(t)| =\left|\left( \int d^3\mathbf{r} \: \phi^*_\uparrow(\bv{r})\phi_\downarrow(\bv{r})/N \right)^N \right|.
\label{cohfact_from_GPE}
\end{align}
We now show that \bref{cohfact_from_GPE} and \bref{decoh_timescale} ought to agree for short times. The evolution of an initially homogeneous condensate mean-field according to \eref{impurityGPE}, for short times and dominant $g_0\left|\psi^{(\alpha)}(\xvec)\right|$ is given by
\begin{align}
\label{GPE_short_time}
\phi_\alpha(\xvec,t)=\sqrt{\rho}\Big(1-\frac{i}{\hbar}g_0\left|\psi^{(\alpha)}(\xvec)\right|^2t\Big).
\end{align}
This is exactly reproduced by the SBM. To see this, we write the mean-field wave function as $\phi(x,t)=\expec{\hat{\Psi}(x,t)}$, where the Bosonic field operator is
\begin{align}
\label{BEC_field_operator_coh}
\hat{\Psi}(\xvec,t)=\sqrt{\rho}+\sum_\qvec\Big[u_\qvec(\xvec)(\tilde{b}_\qvec(t)-\cohstval_\qvec)-v_\qvec^*(\xvec)(\tilde{b}^{\dagger}_\qvec(t)-\cohstval_\qvec^*)\Big],
\end{align}
based on shifted BdG modes for which the initial state $\ket{\cohst}$ is a many-mode coherent state
\begin{align}
\label{BdG_coheren_state_eigen}
\tilde{b}_\qvec\ket{\cohst}=\cohstval_\qvec\ket{\cohst},
\end{align}
with offset $\cohstval_\qvec=\frac{\bar{\kappa}_\qvec}{2\hbar\omega_\qvec}$, using
\begin{align}\label{kappa_bar}
\bar{\kappa}_\qvec=g_0 \sqrt{\rho}  \int d^3\xv  (|\psi^{(p)}({\xvec})|^2 + |\psi^{(s)}({\xvec})|^2 ) [u_\qvec ({\xvec})-v^*_\qvec ({\xvec})].
\end{align} 
It arises through the sudden insertion of the Rydberg impurity at $t=0$ \cite{rammohan_tayloring}.
We then use \eref{BEC_field_operator_coh} and evaluate,
\begin{align}
\label{phi_p_t}
\phi_\alpha(\xvec,t)=\bra{\alpha}\bra{\cohst}\:\:\hat{U}_{ph,\alpha}^{\dagger}(t)\hat{\Psi}(t)\hat{U}_{ph,\alpha}(t)\:\:\ket{\alpha}\ket{\cohst},
\end{align}
where the subscript $\alpha$ implies that we look at the mean field evolving in the presence of a Rydberg impurity in the state $\ket{\alpha}$.
Inserting \bref{U_s_unitary_ope} into \bref{phi_p_t}, exploiting \eref{BdG_coheren_state_eigen} and the usual commutation relations, we can reach for short times
\begin{align}
\label{phi_s_short_time}
\phi_{\alpha}(\xvec,t)&=\phi_0(\xvec)+\sum_\qvec\Bigg[u_\qvec(\xvec)\cohstval_\qvec(-i\omega_\qvec t) -v_\qvec^*(\xvec)\cohstval_\qvec(i\omega_\qvec t)\Bigg]\nonumber \\
&+\alpha\sum_\qvec\frac{\Delta\kappa_\qvec}{\hbar\omega_\qvec}\times\Bigg[u_\qvec(\xvec)(-i\omega_\qvec t)-v_\qvec^*(\xvec)(i\omega_\qvec t)\Bigg].
\end{align}
Converting the sum $\sum_\qvec$ into a continuum integration using $\sum_\qvec \to \int d^3\qvec \frac{{\cal V}}{(2\pi)^3}$ as usual, we can simplify this to
\begin{align}
\label{phi_s_short_time_4}
\phi_{\alpha}&(\xvec,t)=
     \sqrt{\rho} -i t  \frac{g_0}{\hbar} \sqrt{\rho}  \:|\psi^{(\alpha)}(\mathbf{x} )|^2 (\bar{u}_\qvec^2 - \bar{v}^2_\qvec),
\end{align}
which thanks to $\bar{u}_\qvec^2 - \bar{v}^2_\qvec=1$ co-incides with \bref{GPE_short_time}. We have thus shown, that the time-evolution of Bogoliubov operators in the SBM reproduces the same mean-field evolution that we would expect from the GPE as long as the Rydberg impurity is in a specific quantum state. 
We numerically verify \cite{footnote:GPEparameters} the agreement for all relevant times in the inset of \frefp{coherence_factor}{b}, using the high-level language XMDS \cite{xmds:paper,xmds:docu}. The agreement confirms the BdG model underlying \bref{Hspinboson}, since 
coherent processes due to higher order phonon operators are included in the GPE, but to net yet cause a visible deviation.

We have thus established the intuitive picture of coherence tied to the overlap of mean-field wavefunctions and further provided a practical method to evaluate $r(t)$ and a verification of the calculations in \rref{rammohan_tayloring} that underpin \eref{decoh_timescale}. Beyond the present context, \eref{cohfact_from_GPE} can provide a useful way for calculating other internal impurity decoherence times e.g.~for ions \cite{ratschbacher2013decoherence}, as long as impurity BEC interactions are amenable to mean-field theory. We verify in \cite{tiwari:tracking} that they are, and hence the product ansatz for $\ket{\Psi_\alpha}$ should provide a good approximation.

\ssection{Imaging the qubit environment interface}
%
Entanglement between system and environment is responsible for the observed decoherence of the system and encoded in the total state of system and environment.
For our main result, we now show that the Rydberg-BEC realization of a qubit and its environment can provide unique access to the mesoscopically entangled state $\ket{\sub{\Psi}{tot}(t)}$ in \eref{psitot} by inspection of the environment. To make this visible, the experiment must first transform the system-environment entanglement into an entangled state involving only the environment, by bringing the qubit quickly from $\ket{+}$ into $\ket{\downarrow}$ after some variable time $t$ of evolution, and then de-exciting it to a ground-state. This leaves the environment in the mesoscopic superposition $\ket{\sub{\Psi}{ms}(t)}=A[\ket{\sub{\Psi}{$\uparrow$}(t)}+\ket{\sub{\Psi}{$\downarrow$}(t)}]$, where $A=1/\sqrt{2(1+\mbox{Re}[\braket{\sub{\Psi}{$\uparrow$}(t)}{\sub{\Psi}{$\downarrow$}(t)}]}$ normalizes the many-body wave function. This state shows characteristic features in the corresponding total density $\sub{\varrho}{ms}(\bv{r})$. To extract them, we take the difference $\Delta \varrho=\sub{\varrho}{ms}(\bv{r})-\sub{\varrho}{inc}(\bv{r})$ compared to the total density $\sub{\varrho}{inc}(\bv{r})$ one would measure if the Rydberg spin was in a classical mixture and find 
 %
\begin{align}
\Delta\varrho(\bv{r})&=\left(A^2 - \frac{1}{2} \right)(\varrho_\uparrow + \varrho_\downarrow)  \CR
&+ A^2\left(N \frac{ \varphi_\uparrow^*(\bv{r})\varphi_\downarrow(\bv{r})}{\int d^3 \mathbf{x} \varphi_\uparrow^*(\mathbf{x})\varphi_\downarrow(\mathbf{x})} r(t)+ \mbox{c.c.} \right),
\label{dens_diff}
\end{align}
Here, $\varrho_\uparrow$ ($\varrho_\downarrow$) is the total density in the presence of spin $\ket{\uparrow}$ ($\ket{\downarrow}$). Clearly $\Delta\varrho(\bv{r})$ is directly related to $r(t)$. 

We assess the observability of this signal in \fref{coherence_factor}, showing the column density relevant for experiments, which is obtained by integrating \bref{dens_diff} over the $y$-direction, with quantization axis along $z$. In the inset we show the maximum $s(t)$ of $\Delta \varrho({\bf r})$ as red curve. The signal reaches $3\%$ of the bulk density, which should be accessible using high sensitivity density measurements \cite{Gajdacz_cloudsbelowshotnoise_PRL} or electron microscopy \cite{Santra_scanelecBEC_review}. The signal is transient, since from \eref{dens_diff} it must vanish initially and once decoherence is complete. 

Thus a Rydberg qubit embedded in a condensed environment offers unique opportunities to probe the environmental origin of decoherence, through a signal heralding the transient buildup of the mesoscopically entangled state causing decoherence.  Further one could coherently manipulate the initial state of the BEC environment in order to functionalize its impact on the qubit or investigate decoherence free subspaces \cite{Schlosshauer_review_PhysRep} by initialising the qubit in a superposition $(\ket{\nu (l=1,m=1)} +\ket{\nu (l=1,m=-1)})/\sqrt{2}$. 

Being able to image the interface to the environment distinguishes the proposed setup from other open quantum systems where the environment is initially in a thermal state or not all of its degrees of freedom can be imaged simultaneously. At later times, the non-equilibrium dynamics for which we consider only the onset here results in the formation of polarons \cite{nielsen2019critical,skou2021non,PhysRevA.97.022707,camargo2018creation,PhysRevLett.116.105302,camargo2017rydberg}, with many-body correlations that go beyond the methods employed here. Nonetheless also in that case \eref{psitot} holds, and coherence measurements combined with a modified signature $\Delta\varrho(\bv{r})$ can provide an additional experimental handle on polaron formation, not possible for embedded spins in electronic ground-states \cite{Ng_singleatomprobe_PRA,schmidt2018quantum,schmidt2019motional,song2019controlling,ostmann2017cooling,mcendoo2013entanglement,klein2007dynamics,cirone2009collective,bruderer2007polaron,bruderer2008transport,bruderer2006probing,Han_CTP_2020,Zhen_coherence_Twocomp}.

\ssection{Microwave driven Rydberg impurity}
%
To demonstrate that the platform can push the frontier of our understanding how many-body quantum dynamics gives rise to decoherence, we now go beyond the case $\sub{\Omega}{mw}=0$, such that the signal $\Delta\varrho(\bv{r})$ is no longer straightforwardly predictable, but advanced numerical techniques such as bosonic DFT \cite{Benavides_bosonic_DFT_PhysRevLett} or ML-MCTDHB \cite{mlmctdhb_kronke2013, mlmctdhb_pra} would be required and challenged.
By adding a continuous microwave drive, the interaction of BdG excitations with the Rydberg impurity can be explored in more detail. For $\sub{\Omega}{mw}\neq 0$, this seemingly innocent modification of the Hamiltonian has dramatic consequences for the many-body evolution, since the block-structure in the Hamiltonian disappears. Hence approaches used so far are no longer applicable.

\begin{figure}[htb]
 \includegraphics[width=0.99\columnwidth]{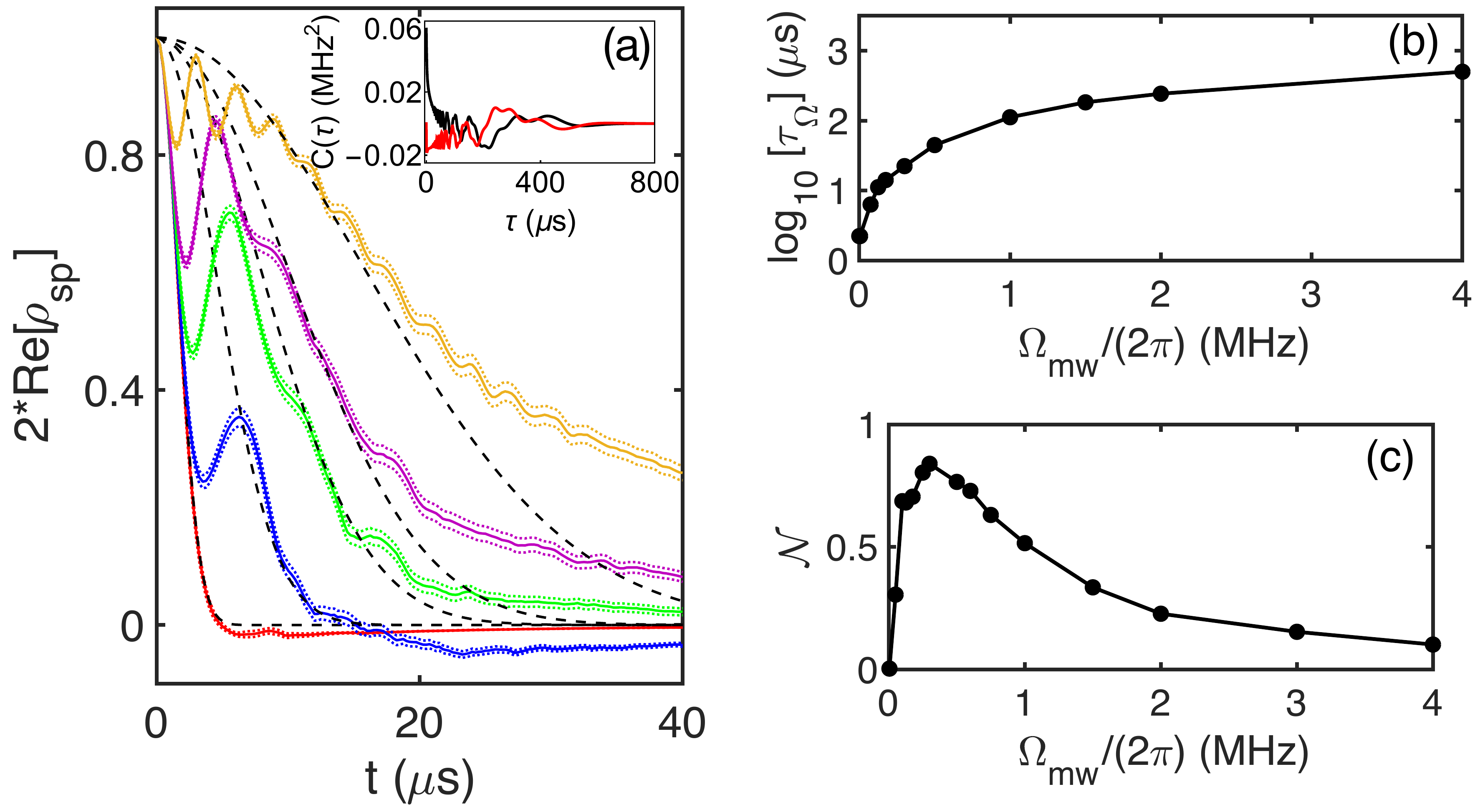}
\caption{\label{fig_non_markovian} Non-Markovian Rydberg decoherence dynamics. (a) 
Coherence between states $\ket{s}$ and $\ket{p}$ at $\nu=120$ as a function of time for increasing resonant ($\Delta=0$) microwave coupling strengths $\sub{\Omega}{mw}$ and initial state $\ket{+}$. From top to bottom $\sub{\Omega}{mw}/(2\pi)=$ $(0.3$, $0.175$, $0.125$, $0.075$, $0.005$) MHz. The black dashed lines show a fit of $\rho_{sp}=\rho_{sp}(0)\exp{[-(t/\tau_\Omega)^2]}$. The inset shows the bath correlation function Re[$C(\tau)$] (black) and Im[$C(\tau)$] (ref) from \eref{bathcorrel_exps} for $\nu=120$, calculated as in \rref{rammohan_tayloring}. (b) Decoherence times $\tau_\Omega$ from fits as in panel (a) as a function of micro-wave strength $\sub{\Omega}{mw}$. (c) Estimated Non-Markovianity ${\cal N}$ based on the net increase of trace distances between selected system states. 
We use 1000 stochastic trajectories. 
}
\end{figure}

For now we focus on the qubit side in this complex scenario only, which permits the use of a computational technique for open quantum systems, namely Non-Markovian quantum state diffusion (NMQSD) \cite{Diosi_NMQSD} solved through the Hierarchy of pure states (HOPS) \cite{Suess_HOPS_PRL}. The method is relatively fast, gives reliable results for the SBM over a large range of parameters \cite{Hartmann_hops_SBM}.
The required system Hamiltonian $\sub{\hat{H}}{syst}$ is given in \bref{Hspinboson}, where now $\Delta$ has become the microwave detuning.
The environmental correlation functions $C(\tau)$ have been determined in \rref{rammohan_tayloring}, and an example is shown in the inset of \frefp{fig_non_markovian}{a}. As required for HOPS, we use the method of \cite{ritschel_specdensfit} to represent the correlation function as a sum of $M$ damped exponentials
\begin{align}
C(\tau)&=\sum_{j=0}^{M-1}  g_j e^{-(i\Omega_j  + \gamma_j) \tau},
\label{bathcorrel_exps}
\end{align}
where $g_j$ is the coupling strength between the system and the environment. The quantum many-body evolution is found through a stochastic wavefunction hierarchy $\psihops^{(\mathbf{k})}$, 
where $\mathbf{k}$ is a vector valued hierarchy index, with one component $k_j$ for each term in the representation of the bath correlation function. Only the zeroth order $\psihops^{(\boldsymbol{0})}=\psihops^{(0,\cdots,0)}$ of the wavefunction hierarchy is used to calculate expectation values of system operators as usual $\expec{\hat{O}}=\bra{\psihops^{(\boldsymbol{0}))}} \hat{O} \ket{\psihops^{(\boldsymbol{0}))}}$. Its time evolution is however coupled to all higher levels according to,
\begin{align}
&\frac{\partial}{\partial t} \psihops^{(\mathbf{k})}(t)=\bigg( -i \sub{\hat{H}}{syst}(t) -\mathbf{k}\cdot \mathbf{w} +  \hat{L}  \tilde{z}_t\bigg)\psihops^{(\mathbf{k})}(t) \CR &+ \sum_{j} k_{j} g_{j} \psihops^{(\mathbf{k}-\mathbf{e}_{j})}(t) -  \sum_{j}   \hat{L}^\dagger  \psihops^{(\mathbf{k}+\mathbf{e}_{j})}(t),
\label{hops_eom}
\end{align}
where $\mathbf{w}=[w_1, \cdots w_M]^T$, with $w_j=i\Omega_j  + \gamma_j$, see \eref{bathcorrel_exps}. Further $\hat{L}$ is the system part of the coupling term to the environment, which is $\hat{L}=\sigma_z$ in Eq. (\ref{Hspinboson}) and $\mathbf{e}_{j}$ is a unit vector along cartesian direction $j$. The system environment coupling enters \bref{hops_eom} twofold, through $\hat{L}=\hat{\sigma}_z$ and through the shifted noise $\tilde{z}_t$, with
\begin{align}
\tilde{z}_t = z^{*}_t  + \int_0^t ds \:\: C^*(t-s )\expec{ \hat{L}^\dagger},
\label{shiftnoise}
\end{align}
where $\langle . \rangle$ denotes the normalized average over $\psihops^{(0)}$.
The unshifted noises $z^{*}_t$ are constructed such that their temporal correlation function is exactly the bath correlation function,
\begin{align}
\overline{z_t  z^{*}_s}=C(t-s),
\label{noisecorrel}
\end{align}
where $\overline{\cdots}$ denotes the stochastic average. 

When the qubit is driven, we find very clear characteristic non-Markovian features in the decoherence dynamics, shown in \frefp{fig_non_markovian}{a}. Instead of a monotonic decay of coherence, there are partial revivals or oscillations. Intuitively, the impurity excites sound waves in the condensate that can impact back on it at a later time \cite{ostmann2017cooling} due to long-range interactions. We extract the overall initial decoherence times $\tau_\Omega$ from Gaussian fits indicated by dashed lines in panel (a) and show them in panel (b). Clearly the decoherence time
is tunable through $\sub{\Omega}{mw}$, in addition to $\nu$ discussed before. Finally, we compare the degree of non-Markovianity ${\cal N}$ for different parameters using the measure developed by Breuer {\it et al.}~\cite{Breuer_NonMarkovMeasure_PRL,genkin:markovswitch}, an estimate of which we show in \frefp{fig_non_markovian}{c}. To quantify the degree of Non-Markovinity, we thus consider two different initial system density matrices $\hat{\rho}=\ket{s}\bra{s}$ and $\hat{\rho}=\ket{p}\bra{p}$, and monitor how the trace distance between the two states evolves in time. Non-Markovinity can be quantified by integrating positive rates of change over time 
\cite{Breuer_NonMarkovMeasure_PRL}. The result is shown in \frefp{fig_non_markovian}{c} It is clear that the non-Markovianity of the system depends on the microwave strength. It was expected that the system behaves non-Markovian, since its characteristic time-scales (for Rydberg systems $\sim \mu$s) are much shorter than the 
the environmental memory time $T_m\approx 650$ $\mu$s for $\nu=120$, set by the relevant phonon frequencies $\omega_\bv{q}\sim 5$ kHz. $T_m$ decreases for smaller $\nu$ and reaches $T_m = 17 \mu$s at $\nu=30$, still is only slightly shorter than the Rydberg life-time, which limits system evolution times. Rydberg qubit decoherence will thus generally remain non-Markovian.
The approach demonstrated in this section also completes the toolkit required for a comprehensive treatment of Rydberg impurities in BEC. For example NMQSD allowed us to verify that phonon induced Rydberg transitions, which would give rise to terms $\sim \hat{\sigma}_y \otimes (\tilde{b}_\mathbf{q}+\tilde{b}_\mathbf{q}^{\dagger})$ in the Hamiltonian \cite{rammohan_tayloring}, are strongly suppressed by the energy mismatch between $\Delta E$ and $\hbar \omega_\qvec$ and hence neglected in \eref{Hspinboson} here.

In the non-Markovian regime just discussed, imaging the qubit to environment interface can give new insights on the back and forth flow of information between the two \cite{Breuer_NonMarkovMeasure_PRL}, provide hints on how to shield qubits from decoherence \cite{Schlosshauer_review_PhysRep} and challenge advanced quantum many body methods \cite{Benavides_bosonic_DFT_PhysRevLett,mlmctdhb_kronke2013}. Even more environment interrogation techniques are available in BEC, such as precision phonon-spectroscopy \cite{Katz_PRL_phononspec}. The measure ${\cal N}$ itself is experimentally accessible when also measuring qubit coherence with Ramsey spectroscopy.

\ssection{Conclusions and outlook} 
%
We propose that the interface between a qubit and its environment can be imaged, if the former is realized by a Rydberg atom and the latter by an embedding BEC.
This opens up an experimental window on both intertwined aspects of decoherence, spin and environment, evolving into a mesoscopic superposition state such as \bref{psitot}.
Experimentally probing the interface through straightforward column densities could test the foundations of open quantum systems, provide insights on the environmental side of decoherence that suggest avenues for its mitigation and benchmark the most advanced numerical many-body techniques. 

This is possible since the Rydberg atom, in a superposition of electronic states, acts as control handle that can affect the BEC environment over an optically resolvable range of micrometers and as a probe that is itself straightforward to read out. Neither of these advantages holds for embedded spins in electronic ground-states \cite{Ng_singleatomprobe_PRA,schmidt2018quantum,schmidt2019motional,song2019controlling,ostmann2017cooling,mcendoo2013entanglement,klein2007dynamics,cirone2009collective,bruderer2007polaron,bruderer2008transport,bruderer2006probing,Han_CTP_2020,Zhen_coherence_Twocomp}.

The proposed experiments can complement tomography of a decohering quantum system \cite{GlKuGu07_297,deleglise:reconstruction}, by interrogating the environmental aspect of decoherence instead.  We demonstrate that the time-scale where the target signal exists can be tuned from the order of nanoseconds to microseconds by choice of the excited Rydberg principal quantum number $\nu$ or additional microwave driving.

\acknowledgments
We gladly acknowledge fruitful discussions with Rick Mukherjee, and thank the Science and Engineering Research Board (SERB), Department of Science and Technology (DST), New Delhi, India, for financial support under research Project No.~EMR/2016/005462, and the Max-Planck society under the MPG-IISER partner group program.  R.N.~acknowledges a UKIERI- UGC Thematic Partnership No.~IND/CONT/G/16-17/73 UKIERI-UGC project
and DST-SERB for the Swarnajayanti fellowship File No.~SB/SJF/2020-21/19. A.E.~acknowledges support from the DFG via a Heisenberg fellowship (Grant No EI 872/5-1).


\begin{thebibliography}{76}
\bibitem[{foo({\natexlab{a}})}]{footnote:GPEparameters}
\bibinfo{note}{We consider a homogeneous condensate with density
  $\rho=9.8\times10^{18}$ m$^{-3}$, $U_0=4\pi\hbar^2 a_s/m$ with scattering
  length $a_s=6.45$ nm and $m=1.393\times 10^{-25}$ kg, $g_0=2\pi\hbar^2
  a_e/m_e$ with electron mass $m_e$, electron-atom scattering length
  $a_e=-18\:a_0$ nm, all corresponding to Strontium $^{84}$Sr.}
\bibitem[{foo({\natexlab{b}})}]{footnote:lifetimes}
\bibinfo{note}{In Sr at density $\rho=9.8\times10^{18}$ m$^{-3}$, we expect
  $\tau\approx 35$ $\mu$s for $\nu=80$ and $\tau\approx 150$ $\mu$s for
  $\nu=120$, based on \cite{Kanungo_Sr_lossrates}.}
\end{thebibliography}

\end{document}